\documentclass[twocolumn,preprintnumbers,amsmath,amssymb]{revtex4}
\usepackage{graphicx}
\usepackage{dcolumn}
\usepackage{bm}

\usepackage{graphicx}  
\begin{document}
\title{
Terahertz and Infrared Photodetection using
 p-i-n Multiple-Graphene-Layer Structures
}
\author{V.~Ryzhii\footnote{Electronic mail: v-ryzhii(at)u-aizu.ac.jp} and 
M.~Ryzhii
}
\address{
Computational Nanoelectronics Laboratory, University of Aizu, 
Aizu-Wakamatsu  965-8580 and Japan Science and Technology Agency, 
CREST, Tokyo 107-0075, Japan
}
\author{V.~Mitin
}
\address{Department of Electrical Engineering,
University at Buffalo, State University of New York, NY 14260, USA
}
\author{T.~Otsuji
}
\address{Research Institute for Electrical Communication, Tohoku University, Sendai 980-8577 and 
Japan Science and Technology Agency, CREST, Tokyo 107-0075, Japan
}
\begin{abstract}
We propose 
to utilize  multiple-graphene-layer 
 structures with lateral p-i-n junctions 
for terahertz (THz) and infrared  (IR)
photodetection
and
substantiate the  operation of photodetectors based on these structures. 
Using the developed device model, we calculate the detector dc responsivity and detectivity as functions of the number of graphene layers and geometrical
parameters and show that the dc responsivity and detectivity
 can be fairly large, particularly,
at the lower end of 
the THz  range at room temperatures.
Due to relatively high quantum efficiency
and low thermogeneration rate,
the photodetectors under consideration can substantially surpass
other  THz and IR detectors.
Calculations of the detector responsivity as a function of modulation
frequency of THz and IR radiation demonstrate that the proposed
photodetectors are very fast and can operate at the modulation frequency 
of several tens of GHz.
\end{abstract}

\maketitle
\newpage

\section{Introduction}
Unique properties of graphene layers (GLs)~\cite{1,2,3} make them promising
for different nanoelectronic device applications. The gapless 
energy spectrum of GLs, which is an obstacle for creating
transistor-based digital circuits, opens up prospects to use GLs in
terahertz (THz) and infrared (IR) devices.
Novel optoelectronic THz and IR devices were proposed and evaluated, in particular, in  Refs.~\cite{4,5,6,7,8,9,10,11}.
Recent success in fabricating  multiple-GL structures
with long momentum relaxation time of electrons and holes~\cite{12}
promises  a significant enhancement of the performance
of futures  graphene
optoelectronic devices~\cite{13}.

In this paper, we study the operation of THz and
IR
photodetectors based on multiple-GL structures with p-i-n  junctions.
We refer to the photodetectors in question as to GL-photodetectors (GLPDs).
We focus on  GLPDs with 
$p$- and $n$-doped sections in GLs near
the side contacts~\cite{14,15},
p$^{+}$ and n$^{+}$ contacts
(for example, made of doped poly-Si~\cite{16}),  
 and  multiple GL-structures with the Ohmic side contacts
and split-gates which provide the formation of the electrically
induced p- and n-sections~\cite{17,18,19,20,21,22}. 
The device structures under considerations
are shown in Figs.~1(a) and 1(b).
It is assumed that the highly-conducting GL(s) between the SiC substrate
and the top GLs is removed.
Multiple GL-structures without this highly conducting
GL can be fabricated using
chemical/mechanical reactions and transferred substrate techniques
(chemically etching the substrate and the highly conducting bottom 
GL~\cite{23} or mechanically 
peeling the  upper GLs, then transferring the upper portion of
the  multiple-GL structure on a Si substrate).

Using the developed  device model, we calculate
the GLPD responsivity and detectivity as THz or IR photodetector
(Secs.~II and III),
evaluate its dynamic response (Sec.~IV), and compare GLPDs with some other
THz and IR photodetectors~(Sec.~V), in particular, with quantum-well infrared photodetectors (QWIPs) and quantum-dot infrared photodetectors (QDIPs).
In Sec.~VI, we discuss possible role of the  Pauli blocking in
the spectral characteristics of GLPDs 
and draw main conclusions.
The calculations related to the effect of  screening
of the vertical electric field
on the formation of the electrically
induced  p- and n-sections in multiple-GL structures are singled out in
the Appendix.
\begin{figure}[t]\label{Fig.1}
\vspace*{-0.4cm}
\begin{center}
\includegraphics[width=6.5cm]{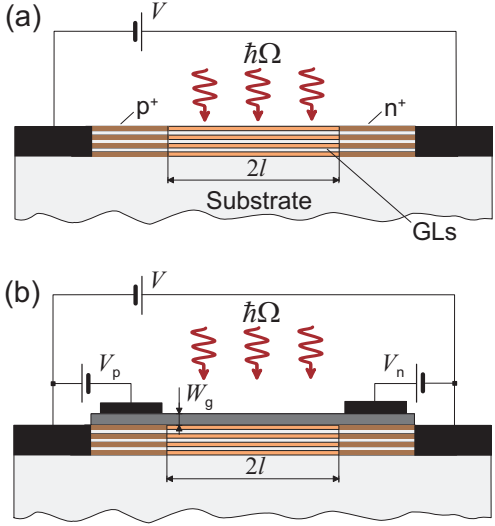}
\caption{Device structures of GLPDs with  (a) doped and (b) electrically
induced $p-i-n$- junctions.
}
\end{center}
\end{figure}

\section{Responsivity and detectivity}

We assume that the intensity of the incident THz or IR radiation 
with the frequency $\Omega$ apart from
the dc component $I_0$ includes 
the ac component:  
$I(t) = I_0 + \delta I_{\omega}\exp(-i\omega)t$,
where $\delta I_{\omega}$ and $\omega$ are the amplitude of the latter component and its modulation frequency, respectively.
In such a situation, the net dc current
(per unit width of the device in the direction perpendicular to
the current)  can be presented in the following form:

\begin{equation}\label{eq1}
J_0 = J_0^{dark} +   J_0^{photo}  
\end{equation}
with
\begin{equation}\label{eq2}
 J_0^{dark} = 4Kel(g_{th}  + g_{tunn})
\end{equation}
and
\begin{equation}\label{eq3}
 J_0^{photo} = 4el\sum_{k = 1}^{K}\frac{\beta_{\Omega}\,I_0^{(k)}}{\hbar\Omega}.
\end{equation}
Here $e$ is the electron charge,
$K$ is the number of GLs, $2l$ is the length of the GL's i-section,
$g_{th}$ and $g_{tunn} = 
(eV/2lv_W^{2/3}\,\hbar)^{3/2}/8\pi^2$
are the  rates of thermal and tunneling 
generation of the electron-hole pairs
(per unit area)~\cite{22},
$V$ is the  bias voltage, $v_W \simeq 10^8$~cm/s is the characteristic velocity of electrons
and holes in graphene, $\beta_{\Omega} = \beta[1 - 2f(\hbar\Omega/2)]$
is the absorption coefficient
of radiation in a GL due to the interband transitions~\cite{24}, where
$\beta = (\pi\,e^2/c\hbar) \simeq 0.023$, 
$f(\varepsilon)$ is the distribution function of electrons and holes in the i-section, 
$\hbar$ is the Planck constant, and $c$ is the speed of light.
The term $g_{tunn}$ in Eq.~(2) is associated with the interband tunneling
in the electric field in the i-section.
The quantity $I_0^{(k)} = I_0(1 - \beta)^{K - k}$ is the intensity of THz
or IR  radiation at the $k$-th GL ($1 \leq k \leq K$).

At a sufficiently strong  reverse
bias, electrons and
holes are effectively swept out from the i-section to the contacts
and heated.
Due to this, one can assume that under the GLPD operation conditions, 
 $f(\hbar\Omega/2) \ll 1$, so that  the distinction between 
$\beta_{\Omega}$
and  $\beta$ can be disregarded (see below).

Using Eqs.~(1) and (3), we arrive at the following formula for
the GLPD dc responsivity:
\begin{equation}\label{eq4}
R_0 = \frac{J_0^{photo}}{2lI_0} =  K^*\frac{2e\beta_{\Omega}}{\hbar\Omega},
\end{equation}
where $K^* = \sum_{k=1}^{K}(1 - \beta)^{K - k} = [1 - (1 - \beta)^K]/\beta$.
Equation~(4) can also be rewritten as
\begin{equation}\label{eq5}
R_0 = [1 - (1 - \beta_{\Omega})^K]\frac{2e}{\hbar\Omega} 
\propto \frac{1}{\hbar\Omega}.
\end{equation}
If $K  = 1$, Eq.~(5) at  $\Omega/2\pi = 1$~THz, yields,
$R_0 \simeq 12$~A/W.
For $K = 50 - 100$ ($K^* \simeq 30 - 39$),
setting $K = 50$, in the frequency range  $\Omega/2\pi = 1 - 10$~THz, 
from Eq.(4) we obtain
$R_0 \simeq 35 - 350$~A/W.

The dark-current limited detectivity, $D^*$,
defined as $D^* = (J^{photo}/NP)\sqrt{A\cdot\Delta f}$, where
$N$ is the noise (in Amperes), $P$ is the power received by the photodetector (in Watts), $A$ is the area of the photodetector (in cm$^2$)(see, for instance,~\cite{25,26}),
and $\Delta f$ is the bandwidth,
can be expressed
via the responsivity $R_0$
and the dark current $J_0^{dark}H$, where
$H$ is the device width in the direction perpendicular 
to the current,    can be presented as
\begin{equation}\label{eq6}
D^* = R_0\sqrt{\frac{A}{4eJ_0^{dark}H}},
\end{equation}
Using Eqs.~(2) - (6), we arrive at

$$
D^* = \frac{K^*}{\sqrt{K}}\frac{\beta_{\Omega}}{\hbar\Omega}
\frac{1}{\sqrt{2(g_{th} + g_{tunn})}}
$$
\begin{equation}\label{eq7}
= \frac{ [1 - (1 - \beta_{\Omega})^K]}{\sqrt{K}\hbar\Omega}
\frac{1}{\sqrt{2(g_{th}  + g_{tunn})}} \propto \frac{1}{\hbar\Omega}.
\end{equation}
Assuming that $K = 1$ and 
$g_{th} = 10^{21}$~cm$^{-2}$s$^{-1}$~\cite{27}, at $T = 300$~K for
$\Omega/2\pi = 1 - 2 $~THz we obtain $D^* \simeq 
(4.1  - 8.2)\times 10^{8}$
cmHz$^{1/2}$/W. Setting $K = 50$, we arrive at
$D^* \simeq (1.7 - 3.4)\times10^{9}$
cm$\cdot$Hz$^{1/2}$/W. 
Due to a significant decrease in the thermogeneration rate at lower temperatures, the detectivity markedly increases with decreasing temperature.
Indeed, at  $T = 77$~K, setting 
$g_{th} = 10^{13}$~cm$^{-2}$s$^{-1}$~\cite{27},
we obtain  $D^*  \simeq (1.7 - 3.4)\times10^{13}$~cm$\cdot$Hz$^{1/2}$/W. 

Figure~2 shows the dependences of the dc responsivity and detectivity 
on the number of GLs $K$ calculated for $\Omega/2\pi = 2$~THz
using Eqs.~(5) and (7) for GLPDs with sufficiently large $l$ in which
the tunneling is weak (see below).
\begin{figure}[t]\label{Fig.2}
\vspace*{-0.4cm}
\begin{center}
\includegraphics[width=7.5cm]{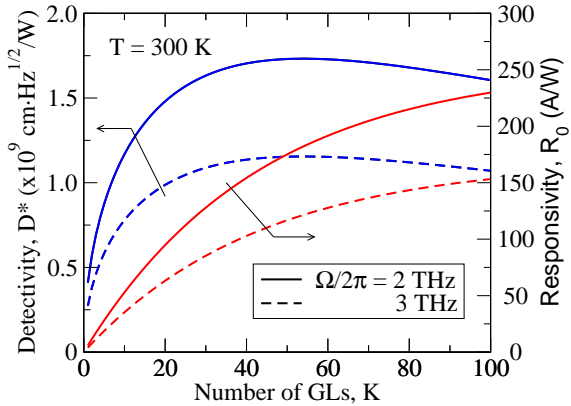}
\caption{Responsivity $R_0$ and detectivity $D^*$ versus number of GLs K
for $\Omega/2\pi = 2$ and 3~THz.
}
\end{center}
\end{figure}

Equation~(7) shows that the GLPD  detectivity decreases with increasing
photon frequency. This is because the spectral dependence of
the GLPD  detectivity is determined by that of the responsivity. 
As follows from Eqs.~(6) and (7),
the detectivity drops at elevated bias voltages when the interband tunneling
prevails over their thermogeneration of the electron and hole pairs.
According to Eqs.~(4) and (7), the GLPD responsivity is independent of
the length of the i-section $2l$ and the bias voltage $V$
(at least if the latter is not too small), whereas 
the detectivity increases  with increasing  $l$ and decreases with
$V$.
This is because the component of the dark current 
associated with the 
thermogeneration rate increases linearly with increasing $l$. 
At the same time,
the tunneling component is proportional to  $V^{3/2}/\sqrt{l}$.
As a result, from Eq.~(7) one can arrive at
\begin{equation}\label{eq8}
D^* \propto \frac{1}{\sqrt{1 + b(V/l)^{3/2}}},
\end{equation}
where
$b \propto 1/g_{th}$. Thus, at elevated bias voltages 
when the tunneling generation surpassed
the thermal generation,  $D^*  \propto V^{-3/4}$.
Figure~3 shows the voltage-dependence of the detectivity $D^*_{tunn}$ with
 the interband tunneling normalized by the detectivity
$D^*$ without tunneling calculated for different values 
of the i-section length $2l$.

Since the  value
 of $D^*$ is determined
by both $l$ and $V$, the latter values should be properly chosen.

There is also a limitation associated 
with the necessity
to satisfy the condition $2l \lesssim l_R$,
where $l_R$ is the recombination length. 
Otherwise, the recombination
in the i-section can become essential resulting in degrading of the GLPD
performance. The recombination length can be defined as
$l_R =  L_V$, where   $L_V \simeq <v>\tau_R$,
$\tau_R$ is the recombination
time,  and  $<v>$ is the average 
drift velocity in the i-region.
 Assuming that for the electron and hole densities
close to that in the intrinsic graphene at $T = 300$~K 
one can put
$\tau_R = 5\times10^{-10}$~s. 
Setting  for sufficiently large bias 
$<v> = 5\times 10^7$~cm/s~\cite{28}, one can find
 $L_V \simeq 250~\mu$m.

\begin{figure}[t]\label{Fig.3}
\vspace*{-0.4cm}
\begin{center}
\includegraphics[width=6.5cm]{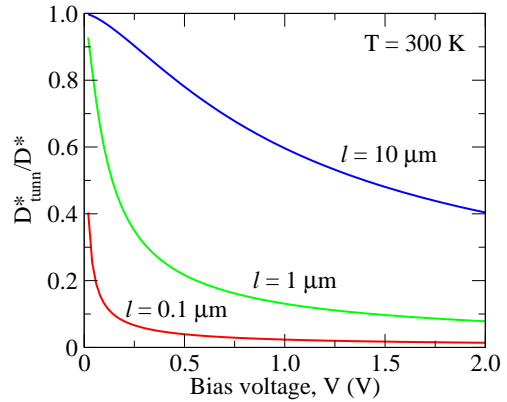}
\caption{Normalized  detectivity versus number bias voltage
for GLDPs with different length of i-section $2l$.
}
\end{center}
\end{figure}
\section{Features of GLPD with electrically-induced  junction}

The main distinctions of the detectors with doped and electrically induced
p-i-n-junction are that the Fermi energies of electrons and holes
in the latter depend on the gate voltages $V_p < 0$ in the p-section
and $V_n > 0$ in the n-section [see Fig.~1(b)]  
and that these energies and, hence, the heights of the barriers
confining electrons and holes in the pertinent section
are different in different GLs. 
This is due to the screening
of the transverse electric field created by the gate voltages by
the GL charges. 
This screening results in a marked decrease in the barrier heights
of the GLs located in the MGL-structure depth (with large indices $k$)
in comparison with
the GLs near the top. In the following, we set $V_n = - V_p = V_g $.
The barriers in question effectively
prevent the electron and hole injection from the contacts
under the reverse bias if the barrier heights are sufficiently  large.
However, the electron and hole injection into the GLs with large $k$
leads to a significant increase in the dark current (in addition to the
current of the electrons and hole generated in the i-section). 
This can substantially
deteriorate the GLPD  detectivity.
As for the responsivity, it can still be calculated using Eqs.~(4) or (5).
To preserve sufficiently high detectivity, the number of GLs should not be too
large.
Indeed, 
taking into account the contributions of the current injected from the
p- and n-regions to the net dark current, the detectivity
 of a GLPD with the electrically-induced
p-i-n junction can be presented as

\begin{equation}\label{eq9}
D^* = \frac{K^*\beta_{\Omega}}{\hbar\Omega\sqrt{2\biggl[\sum_{k=1}^Kj_i^{(k)}/el + K(g_{th} 
+ g_{tunn})\biggr]}},
\end{equation}
where $j_i^{(k)} \propto \exp(- \varepsilon_F^{(k)}/k_BT)$ is the electron  current injected from the p-region
and the hole current injected from the n-region,
$\varepsilon_F^{(k)}$ is the Fermi energy of holes (electrons) in the p-region
(n-region) of the $k$-th GL ($1 \leq k \leq K$).

The injected currents are small in comparison with the current of the 
electrons and holes thermogenerated in the i-region if

\begin{equation}\label{eq10}
\frac{v_W\Sigma_0}{\pi}\exp\biggl(-\frac{\varepsilon_F^{(k)}}{k_BT}\biggr) < 2lg_{th}.
\end{equation}
Here $\Sigma_0$ is the electron and hole density in the intrinsic graphene
(at given temperature).

Considering Eqs.~(A5) and (A6) from the Appendix,
the latter imposes the following limitation on the number of GLs
in GLDs with the electrically-induced p-i-n junctions and the gate voltage $V_g$:
\begin{equation}\label{eq11}
K < K^{max} = \frac{1}{\gamma}\sqrt{\frac{\varepsilon_F^T}{k_BT\ln (v_W\Sigma_0/2\pi\,lg_{th})}} \propto V_g^{1/12}.
\end{equation}
The quantity $\gamma$ is defined in the Appendix.
For $\gamma = 0.07 -0.12$,
assuming that the Fermi energy in the topmost GL $\varepsilon_F^{(1)}
= \varepsilon_F^T = 100$~meV,
 $T = 300$~K, $\Sigma_0 \simeq 8\times10^{10}$~cm$^{-2}$,
$l = 10~\mu$m,
and $g_{th} \simeq 10^{21}$~cm$^{-2}$s$^{-1}$,
we obtain $K^{max} \simeq 34 - 58$, i.e., fairly large.
A decrease in  $l$ results in smaller $K^{max}$ due to a substantial
increase in the tunneling current. For instance, if 
$l = 0.1~\mu$m~\cite{11},
we obtain   $K^{max} \simeq 7 - 10$.

Thus, by  applying  gate voltages one can form the p- and n-sections
with sufficiently high densities and large Fermi energies
of holes and electrons 
in GL-structures with a large number of GLs.
  
\section{Dynamical response}

To describe the dynamic response to the signals  modulated with the frequency
$\omega \ll \Omega$, one needs to find the ac components of the
net density of
photogenerated electron and holes, $\delta\Sigma_{\omega}^{-}$
and $\delta\Sigma_{\omega}^{+}$, respectively. These components are governed
by the following equation:
\begin{equation}\label{eq12}
-i\omega\delta\Sigma_{\omega}^{\mp} 
\pm <v>\frac{d\delta\Sigma_{\omega}^{\mp}}{dx}
= \sum_{k = 1}^K\frac{\beta_{\Omega}\delta\,I_{\omega}^{{k}}}{\hbar\Omega}.
\end{equation}
As in above,
$\delta\,I_{\omega}^{(k)} = \delta\,I_{\omega}(1 - \beta)^{K - k}$,
so that 
$$
\sum_{k = 1}^K\frac{\beta_{\Omega}\delta\,I_{\omega}^{(k)}}{\hbar\Omega}
= K^*\frac{\beta_{\Omega}\delta\,I_{\omega}}{\hbar\Omega}.
$$
In the case of ballistic transport of electrons and holes across the i-section, $<v> \simeq v_W$.
If the electron and hole transport is substantially affected by quasi-elastic
scattering, so that the momentum distributions of electrons and holes are
virtually semi-isotropic, one can put $<v> \simeq v_W/2$ 
(see also, Ref.~\cite{28}).   
Solving Eq.~(12) with the boundary conditions
$\delta\Sigma_{\omega}^{\mp}|_{x = \pm l} = 0$,
we obtain 
\begin{equation}\label{eq13}
\delta\Sigma_{\omega}^{\mp} = 
K^*\frac{\beta_{\Omega}}{\hbar\Omega}
\cdot\frac{\exp[i\omega(l \pm x)/<v>] - 1}{i\omega}\delta\,I_{\omega}.
\end{equation}
Using~Eq.~(13) and considering the Ramo-Shockley
theorem~\cite{28,29} applied to the case of specific contact geometry
~\cite{30}
 the ac current induced in the side contacts (terminal current),
can be presented as 
\begin{equation}\label{eq14}
\frac{\delta\,J_{\omega}^{photo}}{\delta\,I_{\omega}} = K^*\frac{el\beta_{\Omega}}{\pi\hbar\Omega}
\int_{-1}^{1}\frac{d\xi}{\sqrt{1 - \xi^2}}
\frac{[e^{i\omega\tau_t} \cos(\omega\tau_t\xi) - 1]}{i\omega\tau_t}.
\end{equation}
Here we have introduced the characteristic transit time $\tau_t = l/<v>$.
The feature of the contact geometry (the blade-like contacts)
was accounted for by using 
the form-factor $g(\xi) = 1/\pi\sqrt{1 - \xi^2}$~\cite{31}.
This is valid because the thickness of the side contacts and
multiple-GL structure under consideration $Kd \ll l$ 
even at rather large numbers
of GLs $K$, where $d$ is the spacing between GLs.
Similar approach was used previously to analyze the dynamic
response of the lateral $p-n$
junction photodiodes made of the standard semiconductors~\cite{32}
and the graphene tunneling transit-time THz oscillator~\cite{22}.
Integrating in Eq.~(14), we arrive at the following:

$$
\frac{\delta\,J_{\omega}^{photo}}{\delta\,I_{\omega}} =  K^*\frac{4el\beta_{\Omega}}
{\hbar\Omega}
\biggl[\frac{\sin(\omega\tau_t){\cal J}_0(\omega\tau_t)}{\omega\tau_t}
$$
\begin{equation}\label{eq15}
+ i \frac{1 - \cos(\omega\tau_t){\cal J}_0(\omega\tau_t)}{\omega\tau_t}
\biggr],
\end{equation}
where ${\cal J}_0(\xi)$ is the Bessel function. 
Equation~(15) yields 
\begin{equation}\label{eq16}
\frac{|\delta\,J_{\omega}^{photo}|}{\delta\,I_{\omega}}
 =  K^*\frac{4el\beta_{\Omega}}{\hbar\Omega}
\frac{\sqrt{1 - 2\cos(\omega\tau_t){\cal J}_0(\omega\tau_t) +
{\cal J}^2_0(\omega\tau_t)} }{\omega\tau_t}.
\end{equation}
Using Eq.~(16), for the frequency dependent responsivity $R_{\omega} 
= |\delta\,J_{\omega}^{photo}|/2l\delta\,I_{\omega}$,
we obtain
$$
R_{\omega} =  K^*\frac{2e\beta}{\hbar\Omega}
\frac{\sqrt{1 - 2\cos(\omega\tau_t){\cal J}_0(\omega\tau_t)+
{\cal J}^2_0(\omega\tau_t)}}{\omega\tau_t}
$$
\begin{equation}\label{eq17}
= R_0\frac{\sqrt{1 - 2\cos(\omega\tau_t){\cal J}_0(\omega\tau_t)+
{\cal J}^2_0(\omega\tau_t)}}{\omega\tau_t}.
\end{equation}
\begin{figure}[t]\label{Fig.3}
\vspace*{-0.4cm}
\begin{center}
\includegraphics[width=6.5cm]{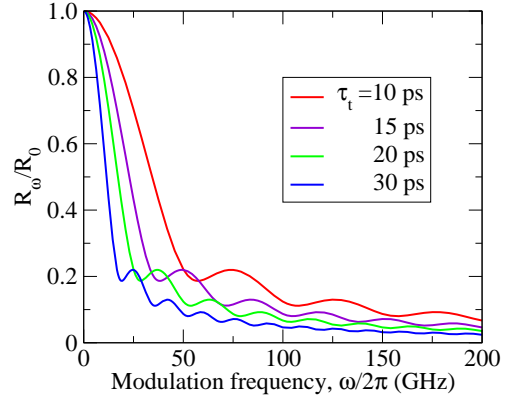}
\caption{Dependence of responsivity (normalized) on 
modulation frequency $\omega/2\pi$
for different $\tau_t$.
}
\end{center}
\end{figure} 

Figure~3 shows the
  responsivity $R_{\omega}$ normalized by its dc value
versus the modulation frequency $\omega$ calculated using Eq.~(17)
for different values of the transit time $\tau_t \propto l$.
At $<v> = v_W/2 \simeq 5\times 10^7$~cm/s, the range $\tau_t = 10 - 30$~ps
corresponds to the length of the i-section $2l = 10 - 30~\mu$m.
As seen from Fig.~3, $R_{\omega}/R_0 = 1/\sqrt{2}$ when the modulation
cut-off frequency $f_t  = \omega_t/2\pi 
\simeq 8.3 - 24.9$~GHz. Naturally,
in GLPDs with shorter i-sections, $f_t$ can be markedly larger (although, at the expense of a substantial increase in the
tunneling current).
Due to this, GLPDs can be used as ultrafast THz and  IR photodetectors.

\section{Comparison with QWIPs and QDIPs}

Now we compare the responsivity and detectivity of GLPDs 
with $K$ GLs calculated above
and those of QWIPs (properly coupled with the incident THz or IR radiation)
 and QDIPs
with the same number of QWs. 
The fraction of the absorbed photon flux in one QW $\beta_{\Omega}^{(QW)} \simeq 
\sigma_{\Omega}^{(QW)}\Sigma_0^{(QW)}$, where $\sigma^{(QW)}$ is the cross-section of the photon
absorption due to the intersubband transitions and $\Sigma_d$
the donor sheet density. Setting the usual values 
$\sigma_{\Omega}^{(QW)} \simeq 2\times10^{-15}$~cm$^2$ and 
$\Sigma_O^{(QW)} \simeq 10^{12}$~cm$^{-2}$, one obtains 
$\beta_{\Omega}^{(QW)} \simeq 0.002$.
This value is one order of magnitude smaller that $\beta_{\Omega}$.
Hence, one can neglect the attenuation of radiation in QWIPs with 
$K \lesssim 100$ (whereas in GLPDs it can be essential) .
In such a case,
the responsivity of QWIPs
is independent of the number of QWs $K$ in the QWIP structure~\cite{33} 
and given by
\begin{equation}\label{eq18}
R_0^{(QW)} = \frac{e\beta_{\Omega}^{(QW)}}{\hbar\Omega\, p_c},
\end{equation}
where $p_c$ is the so-called capture probability which
relates to the QWIP gain $G^{(QW)}$ as $G^{(QW)} = (1 - p_c)/K p_c \simeq
1/K p_c$~\cite{33}.
Using Eqs.~(4) and (18), we obtain
\begin{equation}\label{eq19}
\frac{R_0}{R_0^{(QW)}} \simeq \frac{2\beta_{\Omega}\,p_c}{\beta_{\Omega}^{(QW)}}K^*.  
\end{equation}
Assuming that  $K = 50 - 100$,  ($K^* \simeq 30 - 39$),  
$\beta_{\Omega}^{(QW)} = 0.002$ and $p_c =0.1$,
we obtain $R_0/R_0^{(QW)} \simeq 70 - 90$.

The ratio of the detectivities can be presented as 

\begin{equation}\label{eq20}
\frac{D^*}
{D^{*(QW)}} \simeq
 \frac{2\beta_{\Omega}\sqrt{p_c}}{\beta_{\Omega}^{(QW)}}
\frac{K^*}{K}
\sqrt{\frac{g_{th}^{(QW)}}{g_{th}}}.  
\end{equation}
As can be extracted  from Ref.~\cite{26},
 $g_{th} \propto \exp(-\hbar\omega_0/k_BT)$~, 
where $\hbar\omega_0 \simeq 0.02$~eV is the energy of optical phonon in GLs, 
whereas  
$g_{th}^{(QW)} \propto \exp(- \varepsilon^{(QW)}/k_BT)$, 
where $\varepsilon^{(QW)}$
is the QW ionization  energy ($\hbar\Omega \gtrsim \varepsilon^{(QW)}$),
 Eq.~(17) yields

\begin{equation}\label{eq21}
\frac{D^*}
{D^{*(QW)}} \propto
 \frac{2\beta_{\Omega}\sqrt{p_c}}{\beta_{\Omega}^{(QW)}}
\frac{K^*}{K}\exp\biggl[\frac{\hbar(\omega_0 - \Omega)}{2k_BT}\biggr].  
\end{equation}
Different dependences of the responsivities and detectivities
of GLDs and QWIPs on $K$ is due different directions of the dark current
and photocurrent: parallel to the GL plane in the former case and 
perpendicular
to the QW plane in the latter case.
The product of the factors in the right-hand side of Eq.~(21) except
the last one can be  on the order of unity. This is because a large
ratio $2\beta_{\Omega}/\beta_{\Omega}^{(QW)}$ can be compensated by relatively small
capture parameter $p_c$. However, the exponential factor in Eq.~(21)
is large in the THz range: $\Omega \lesssim \omega_0$,
i.e., at $\Omega/2\pi \lesssim 50$~THz. 
In particular, at $T = 300$~K and $\Omega/2\pi = 5 - 10$~THz,
the exponential factor in question is about 24 - 36.

The GLPD responsivity and detectivity can also 
markedly exceed those of QDIPs (for which formulas similar to Eqs.~(21) 
and (22) can be used) 
despite lower capture probability and thermoexcitation rate
in QDIPs in comparison with QWIPs
(see, for instance, Refs.~\cite{34,35,36}.
The ratios $D^*/D^{*(QW)}$ 
and $D^*/D^{*(QD)}$ dramatically increases with decreasing $\hbar\Omega$
and $T$. This is attributed to the fact that the spectral dependence
of the GLPDs detectivity is similar to the spectral
dependence of the responsivity ($D^* \propto \beta_{\Omega}/\hbar\Omega$), whereas the GLPDs intended for the photodetection in different spectral ranges exhibit the same dark current (due to the gapless energy spectrum).
In contrast, in QWIPs, QDIPs, and some other photodetectors
(see, for example,~Refs.~\cite{9,10})
the transition to lower photon frequency requires to utilize 
the structures with lower ionization energy and, hence,
exponentially higher dark current. 
The latter leads to quite different spectral dependencies of
the GLPD detectivity and the detectivity of QWIPs and QDIPs 
(which drops with
increasing photon freqiency).

GLPDs can   surpuss QWIPs and QDIPs in responsivity even at relatively
high photon frequencies.
For example, for a GLPD with $K = 25$ 
at
$\Omega/2\pi = 75 $~THz (wavelength $\lambda$ about $4~\mu$m), 
we arrive at $R_0 \simeq 3$~A/W.  This value is
three times  larger 
than $R_0^{(QD)}$ obtained experimentally for a QDIP  with 25
InAs QD layers~\cite{37}. 
As for the detectivity, 
comparing a GLPD with $K = 70$  operating at  $T = 300$~K and 
$\Omega/2\pi \simeq 15$~THz
($\lambda \simeq  20~\mu$m) and a QDIP with 70 QD layers ~\cite{38}, we 
obtain $D^* \sim 2\times 10^{8}$~cm$
\cdot$Hz$^{1/2}$/W and $D^{*(QD)} 
\sim
 10^7$~cm$\cdot$Hz$^{1/2}$/W, respectively.

GLPDs can surpass  also photodetectors
on narrow-gap  and gapless bulk semiconductors
like HgCdTe (BSPDs). Apart from advantages associated
with potentially simpler fabrication, GLPDs might exhibit higher
detectivity (compare  the data above and those from Ref.~\cite{39}).
This can be attributed to relatively low  thermogeneration rate in GLPDs
compared to  BSPDs with very narrow or zeroth gap.  
The point is that thermogeneration rate in GLPDs 
at room temperatures  is primarily due absorption 
of optical phonons~\cite{27}
 which
have fairly large energy $\hbar\omega_0$ (the Auger processes 
in GLs are forbidden~\cite{40}), whereas this rate in  BSPDs
is essentially determined by the Auger processes which are strong~\cite{39}.
 
\section{Discussion and conclusions}

Analyzing Eqs.~(4) - (7), we accepted that $f(\hbar\Omega) \ll 1$
and, therefore,  disregarded the frequency dependence of
the absorption coefficient $\beta_{\Omega}$ associated with the population
of the low energy states by electrons and holes.
For a rough estimate, the value of distribution function $f(0)$
can be presented as $f(0) \lesssim f_0(0)\Sigma/\Sigma_0$, where
$\Sigma$ is the electron and hole density in the i-section
under the reverse bias. 
This density, in turn, can be estimated as
$\Sigma = 2g_{th}\tau_t = g_{th}2l/<v>$. Using the same parameters
as above, for $l = 10~\mu$m   and $T = 30$~K, we obtain
$\Sigma/\Sigma_0 = 0.25$, so that $f(0) \lesssim 0.125$ and
$\beta_{\Omega}|_{\Omega \rightarrow 0} \gtrsim 0.75\beta$.
It implies that the numerical data obtained above for the frequencies 
at the lower end of the THz range might be  slightly overestimated.
However, since the electron-hole system in the i-section can be
pronouncedly
heated by the electric field, the actual values of $f(\hbar\Omega/2)$
can be smaller than in the latter estimate and, hence, $\beta_{\Omega}$
can
be rather close to $\beta$. The electron and hole heating
in  intrinsic GLs under the electric field was studied recently~\cite{28,41}.
In contrast to the cases considered in Refs.~\cite{28,41},
the finiteness of the transit time of electrons and holes in the i-section
of the GL-structures under consideration
strongly affects the electron and hole heating. Therefore, more careful
calculation of $\beta_{\Omega}$ at relatively low $\Omega$ requires separate
consideration.

In summary,
we proposed and evaluated GLPDs   multiple-GL
p-i-n structures. It was shown that GLPDs can exhibit high responsivity
and detectivity in the THz and IR ranges at room temperatures.
Due to relatively high quantum efficiency
and low thermogeneration rate,
the GLPD responsivity and detectivity   can substantially exceed those of
 other photodetectors.

\section*{Acknowledgments}
The authors are grateful to  M.~S.~Shur, A.~A.~Dubinov, V.~V.~Popov,
 A.~Satou, M.~Suemitsu, and F.~T.~Vasko
for fruitful discussions and comments.
This work was supported by the Japan Science and Technology Agency, CREST, 
Japan.

\section*{Appendix. Effect of vertical  screening in multiple GL-structures}
\setcounter{equation}{0}
\renewcommand{\theequation} {A\arabic{equation}}

As was pointed out above,
 the thickness of  multiple GL-structures even with rather large number of GLs $K$
is in reality small in comparison with the lateral sizes of the device, 
namely, the lengths of all the section and, hence, the  
gates. Owing to this,
the distribution of the dc electric potential 
$\psi = 2\varphi_0/V_g$ normalized by $V_g/2$  (where
$V_g = V_n = -V_p$) 
in the direction perpendicular
to the GL plane (corresponding to the axis $z$) can be found from
the one-dimensional Poisson equation:  
\begin{equation}\label{A1}
\frac{d^2\psi}{dz^2} = \frac{8\pi\,e}{\ae\,V_g}\sum_{k=1}^K\Sigma^{(k)}
\delta(z - kd +d).
\end{equation}
Here $\Sigma_0^{(k)}$ is the electron (hole)
density in the $k$-th GL in the n-section (p-section),
 $d$ is the spacing between GLs, and $\delta(z)$ is the Dirac delta-function.
Considering that $\Sigma_0^{(k)} = (\varepsilon_F^{{k}})^2/\pi\hbar^2v_F^2
= (e^2V_g^2/4\pi\hbar^2v_F^2)\,\psi^2|_{z = kd}$, where  $\varepsilon_F^{{k}}$
is the Fermi energy in the $n$-section ($p$-section)
 of the $k$-th GL,
and replacing the summation in Eq.~(A1)
by integration (that is valid if $K$ is not too small), we reduce Eq.~(A1) to the following:
\begin{equation}\label{A2}
\frac{d^2\psi}{dz^2} = \frac{\psi^2}{L_s^2}
\end{equation}
with the characteristic screening length
$L_s = \hbar\,v_F\sqrt{\ae\,d/2e^3V_g} \propto V_g^{-1/2}$.
One can assume that $\psi|_{z = 0} = 1 + 2W_g
(d\psi/dz)|_{z = 0} $ and  $\psi|_{z = \infty} = 0$
(as well as $(d\psi/dz)|_{z = \infty} = 0$),
where $W_g$ is the thickness of the layer separating the multiple GL-structure
 and the gates.
Solving Eq.~(A2) with the latter boundary conditions,
we arrive at
\begin{equation}\label{A3}
\psi = \frac{1}{(C + z/\sqrt{6}L_s)^2},
\end{equation}
where $C$ satisfies the following equation:
\begin{equation}\label{A4}
C^3 - C^2 = (4W_g/\sqrt{6}L_s), 
\end{equation}
Since in reality $4W_g \gg \sqrt{6}L_s$, from Eq.~(A4) one obtains 
$C \simeq (4W_g/\sqrt{6}L_s)^{1/3} \propto V_g^{1/6}$
Setting $d = 0.35$~nm, $W_g = 10$~nm, $\ae = 4$,  and $V_g = 2$~V, one can
obtain $L_s \simeq 0.3$~nm,  $(4W_g/\sqrt{6}L_s) \simeq 53$ and 
$C \simeq 3.75$.

Equation~(A3) yields
\begin{equation}\label{A5}
\varepsilon_F^{(k)} \simeq  \frac{eV_g}{2[C + (k - 1)d/\sqrt{6}L_s]^2}
= \varepsilon_F^T\,a^{(k)}.
\end{equation}
Here $\varepsilon_F^T = \varepsilon_F^{(1)} = eV_g/2C \propto V_g^{5/6}$ 
is the Fermi energy of electrons
in the topmost GL in the n-section  (holes in the p-section) and 
\begin{equation}\label{A6}
a^{(k)} = [1 + (k - 1)\gamma]^{-2},
\end{equation}
where 
$\gamma = d/\sqrt{6}L_sC \propto V_g^{1/3}$
At the above parameters (in particular, $W_g = 10 - 50$~nm and $V_g = 2$V)
$\gamma = d/\sqrt{6}L_sC \simeq 0.07 - 0.12$.



\end{document}